\documentclass[a4paper,conference]{IEEEtran}

\IEEEsettopmargin{t}{30mm}
\IEEEquantizetextheight{c}
\IEEEsettextwidth{14mm}{14mm}
\IEEEsetsidemargin{c}{0mm}

\usepackage[utf8]{inputenc}
\usepackage[T1]{fontenc}
\usepackage{url}
\usepackage{ifthen}
\usepackage{cite}
\usepackage[cmex10]{amsmath} 
\usepackage{graphicx}
\usepackage{amssymb}
\usepackage{amsthm}
\usepackage{subfigure}
\usepackage{booktabs} 
\usepackage{multirow}
\usepackage{microtype}
\usepackage{balance}
\usepackage{xcolor}
\usepackage{xfrac}    
\usepackage{algorithm}
\usepackage{setspace}
\usepackage{mdframed}
\usepackage{tikz}
\usepackage{circledsteps}
\usepackage{algorithmicx}
\usepackage{algpseudocode}
\algrenewcommand\algorithmicindent{0.7em}%
\usepackage{xpatch}

\usepackage{amssymb}
\usepackage{amsfonts}
\usepackage{mathrsfs}
\usepackage{xspace}
\usepackage{bm}
\usepackage{upgreek}

\newcommand{\safemath}[2]{\newcommand{#1}{\ensuremath{#2}\xspace}}

\safemath{\bma}{\mathbf{a}}
\safemath{\bmb}{\mathbf{b}}
\safemath{\bmc}{\mathbf{c}}
\safemath{\bmd}{\mathbf{d}}
\safemath{\bme}{\mathbf{e}}
\safemath{\bmf}{\mathbf{f}}
\safemath{\bmg}{\mathbf{g}}
\safemath{\bmh}{\mathbf{h}}
\safemath{\bmi}{\mathbf{i}}
\safemath{\bmj}{\mathbf{j}}
\safemath{\bmk}{\mathbf{k}}
\safemath{\bml}{\mathbf{l}}
\safemath{\bmm}{\mathbf{m}}
\safemath{\bmn}{\mathbf{n}}
\safemath{\bmo}{\mathbf{o}}
\safemath{\bmp}{\mathbf{p}}
\safemath{\bmq}{\mathbf{q}}
\safemath{\bmr}{\mathbf{r}}
\safemath{\bms}{\mathbf{s}}
\safemath{\bmt}{\mathbf{t}}
\safemath{\bmu}{\mathbf{u}}
\safemath{\bmv}{\mathbf{v}}
\safemath{\bmw}{\mathbf{w}}
\safemath{\bmx}{\mathbf{x}}
\safemath{\bmy}{\mathbf{y}}
\safemath{\bmz}{\mathbf{z}}
\safemath{\bmzero}{\mathbf{0}}
\safemath{\bmone}{\mathbf{1}}

\bmdefine{\biad}{a}
\bmdefine{\bibd}{b}
\bmdefine{\bicd}{c}
\bmdefine{\bidd}{d}
\bmdefine{\bied}{e}
\bmdefine{\bifd}{f}
\bmdefine{\bigd}{g}
\bmdefine{\bihd}{h}
\bmdefine{\biid}{i}
\bmdefine{\bijd}{j}
\bmdefine{\bikd}{k}
\bmdefine{\bild}{l}
\bmdefine{\bimd}{m}
\bmdefine{\bind}{n}
\bmdefine{\biod}{o}
\bmdefine{\bipd}{p}
\bmdefine{\biqd}{q}
\bmdefine{\bird}{r}
\bmdefine{\bisd}{s}
\bmdefine{\bitd}{t}
\bmdefine{\biud}{u}
\bmdefine{\bivd}{v}
\bmdefine{\biwd}{w}
\bmdefine{\bixd}{x}
\bmdefine{\biyd}{y}
\bmdefine{\bizd}{z}

\bmdefine{\bixid}{\xi}
\bmdefine{\bilambdad}{\lambda}
\bmdefine{\bimud}{\mu}
\bmdefine{\bithetad}{\theta}
\bmdefine{\biphid}{\phi}
\bmdefine{\bideltad}{\delta}

\safemath{\bmia}{\biad}
\safemath{\bmib}{\bibd}
\safemath{\bmic}{\bicd}
\safemath{\bmid}{\bidd}
\safemath{\bmie}{\bied}
\safemath{\bmif}{\bifd}
\safemath{\bmig}{\bigd}
\safemath{\bmih}{\bihd}
\safemath{\bmii}{\biid}
\safemath{\bmij}{\bijd}
\safemath{\bmik}{\bikd}
\safemath{\bmil}{\bild}
\safemath{\bmim}{\bimd}
\safemath{\bmin}{\bind}
\safemath{\bmio}{\biod}
\safemath{\bmip}{\bipd}
\safemath{\bmiq}{\biqd}
\safemath{\bmir}{\bird}
\safemath{\bmis}{\bisd}
\safemath{\bmit}{\bitd}
\safemath{\bmiu}{\biud}
\safemath{\bmiv}{\bivd}
\safemath{\bmiw}{\biwd}
\safemath{\bmix}{\bixd}
\safemath{\bmiy}{\biyd}
\safemath{\bmiz}{\bizd}

\safemath{\bmxi}{\bixid}
\safemath{\bmlambda}{\bilambdad}
\safemath{\bmmu}{\bimud}
\safemath{\bmtheta}{\bithetad}
\safemath{\bmphi}{\biphid}
\safemath{\bmdelta}{\bideltad}

\safemath{\bA}{\mathbf{A}}
\safemath{\bB}{\mathbf{B}}
\safemath{\bC}{\mathbf{C}}
\safemath{\bD}{\mathbf{D}}
\safemath{\bE}{\mathbf{E}}
\safemath{\bF}{\mathbf{F}}
\safemath{\bG}{\mathbf{G}}
\safemath{\bH}{\mathbf{H}}
\safemath{\bI}{\mathbf{I}}
\safemath{\bJ}{\mathbf{J}}
\safemath{\bK}{\mathbf{K}}
\safemath{\bL}{\mathbf{L}}
\safemath{\bM}{\mathbf{M}}
\safemath{\bN}{\mathbf{N}}
\safemath{\bO}{\mathbf{O}}
\safemath{\bP}{\mathbf{P}}
\safemath{\bQ}{\mathbf{Q}}
\safemath{\bR}{\mathbf{R}}
\safemath{\bS}{\mathbf{S}}
\safemath{\bT}{\mathbf{T}}
\safemath{\bU}{\mathbf{U}}
\safemath{\bV}{\mathbf{V}}
\safemath{\bW}{\mathbf{W}}
\safemath{\bX}{\mathbf{X}}
\safemath{\bY}{\mathbf{Y}}
\safemath{\bZ}{\mathbf{Z}}

\safemath{\bZero}{\mathbf{0}}
\safemath{\bOne}{\mathbf{1}}
\safemath{\bDelta}{\mathbf{\Delta}}
\safemath{\bLambda}{\mathbf{\UpLambda}}
\safemath{\bPhi}{\mathbf{\Upphi}}
\safemath{\bSigma}{\mathbf{\Upsigma}}
\safemath{\bOmega}{\mathbf{\Upomega}}
\safemath{\bTheta}{\mathbf{\Uptheta}}

\bmdefine{\biAd}{A}
\bmdefine{\biBd}{B}
\bmdefine{\biCd}{C}
\bmdefine{\biDd}{D}
\bmdefine{\biEd}{E}
\bmdefine{\biFd}{F}
\bmdefine{\biGd}{G}
\bmdefine{\biHd}{H}
\bmdefine{\biId}{I}
\bmdefine{\biJd}{J}
\bmdefine{\biKd}{K}
\bmdefine{\biLd}{L}
\bmdefine{\biMd}{M}
\bmdefine{\biOd}{N}
\bmdefine{\biPd}{O}
\bmdefine{\biQd}{P}
\bmdefine{\biRd}{R}
\bmdefine{\biSd}{S}
\bmdefine{\biTd}{T}
\bmdefine{\biUd}{U}
\bmdefine{\biVd}{V}
\bmdefine{\biWd}{W}
\bmdefine{\biXd}{X}
\bmdefine{\biYd}{Y}
\bmdefine{\biZd}{Z}

\bmdefine{\biDelta}{\Delta}
\bmdefine{\biLambda}{\Lambda}
\bmdefine{\biPhi}{\Phi}
\bmdefine{\biSigma}{\Sigma}
\bmdefine{\biOmega}{\Omega}
\bmdefine{\biTheta}{\Theta}

\safemath{\bimA}{\biAd}
\safemath{\bimB}{\biBd}
\safemath{\bimC}{\biCd}
\safemath{\bimD}{\biDd}
\safemath{\bimE}{\biEd}
\safemath{\bimF}{\biFd}
\safemath{\bimG}{\biGd}
\safemath{\bimH}{\biHd}
\safemath{\bimI}{\biId}
\safemath{\bimJ}{\biJd}
\safemath{\bimK}{\biKd}
\safemath{\bimL}{\biLd}
\safemath{\bimM}{\biMd}
\safemath{\bimN}{\biNd}
\safemath{\bimO}{\biOd}
\safemath{\bimP}{\biPd}
\safemath{\bimQ}{\biQd}
\safemath{\bimR}{\biRd}
\safemath{\bimS}{\biSd}
\safemath{\bimT}{\biTd}
\safemath{\bimU}{\biUd}
\safemath{\bimV}{\biVd}
\safemath{\bimW}{\biWd}
\safemath{\bimX}{\biXd}
\safemath{\bimY}{\biYd}
\safemath{\bimZ}{\biZd}

\safemath{\bimDelta}{\biDelta}
\safemath{\bimLambda}{\biLambda}
\safemath{\bimPhi}{\biPhi}
\safemath{\bimSigma}{\biSigma}
\safemath{\bimOmega}{\biOmega}
\safemath{\bimTheta}{\biTheta}

\safemath{\setA}{\mathcal{A}}
\safemath{\setB}{\mathcal{B}}
\safemath{\setC}{\mathcal{C}}
\safemath{\setD}{\mathcal{D}}
\safemath{\setE}{\mathcal{E}}
\safemath{\setF}{\mathcal{F}}
\safemath{\setG}{\mathcal{G}}
\safemath{\setH}{\mathcal{H}}
\safemath{\setI}{\mathcal{I}}
\safemath{\setJ}{\mathcal{J}}
\safemath{\setK}{\mathcal{K}}
\safemath{\setL}{\mathcal{L}}
\safemath{\setM}{\mathcal{M}}
\safemath{\setN}{\mathcal{N}}
\safemath{\setO}{\mathcal{O}}
\safemath{\setP}{\mathcal{P}}
\safemath{\setQ}{\mathcal{Q}}
\safemath{\setR}{\mathcal{R}}
\safemath{\setS}{\mathcal{S}}
\safemath{\setT}{\mathcal{T}}
\safemath{\setU}{\mathcal{U}}
\safemath{\setV}{\mathcal{V}}
\safemath{\setW}{\mathcal{W}}
\safemath{\setX}{\mathcal{X}}
\safemath{\setY}{\mathcal{Y}}
\safemath{\setZ}{\mathcal{Z}}
\safemath{\emptySet}{\varnothing}

\safemath{\colA}{\mathscr{A}}
\safemath{\colB}{\mathscr{B}}
\safemath{\colC}{\mathscr{C}}
\safemath{\colD}{\mathscr{D}}
\safemath{\colE}{\mathscr{E}}
\safemath{\colF}{\mathscr{F}}
\safemath{\colG}{\mathscr{G}}
\safemath{\colH}{\mathscr{H}}
\safemath{\colI}{\mathscr{I}}
\safemath{\colJ}{\mathscr{J}}
\safemath{\colK}{\mathscr{K}}
\safemath{\colL}{\mathscr{L}}
\safemath{\colM}{\mathscr{M}}
\safemath{\colN}{\mathscr{N}}
\safemath{\colO}{\mathscr{O}}
\safemath{\colP}{\mathscr{P}}
\safemath{\colQ}{\mathscr{Q}}
\safemath{\colR}{\mathscr{R}}
\safemath{\colS}{\mathscr{S}}
\safemath{\colT}{\mathscr{T}}
\safemath{\colU}{\mathscr{U}}
\safemath{\colV}{\mathscr{V}}
\safemath{\colW}{\mathscr{W}}
\safemath{\colX}{\mathscr{X}}
\safemath{\colY}{\mathscr{Y}}
\safemath{\colZ}{\mathscr{Z}}

\safemath{\opA}{\mathbb{A}}
\safemath{\opB}{\mathbb{B}}
\safemath{\opC}{\mathbb{C}}
\safemath{\opD}{\mathbb{D}}
\safemath{\opE}{\mathbb{E}}
\safemath{\opF}{\mathbb{F}}
\safemath{\opG}{\mathbb{G}}
\safemath{\opH}{\mathbb{H}}
\safemath{\opI}{\mathbb{I}}
\safemath{\opJ}{\mathbb{J}}
\safemath{\opK}{\mathbb{K}}
\safemath{\opL}{\mathbb{L}}
\safemath{\opM}{\mathbb{M}}
\safemath{\opN}{\mathbb{N}}
\safemath{\opO}{\mathbb{O}}
\safemath{\opP}{\mathbb{P}}
\safemath{\opQ}{\mathbb{Q}}
\safemath{\opR}{\mathbb{R}}
\safemath{\opS}{\mathbb{S}}
\safemath{\opT}{\mathbb{T}}
\safemath{\opU}{\mathbb{U}}
\safemath{\opV}{\mathbb{V}}
\safemath{\opW}{\mathbb{W}}
\safemath{\opX}{\mathbb{X}}
\safemath{\opY}{\mathbb{Y}}
\safemath{\opZ}{\mathbb{Z}}
\safemath{\opZero}{\mathbb{O}}
\safemath{\identityop}{\opI}

\safemath{\veca}{\bma}
\safemath{\vecb}{\bmb}
\safemath{\vecc}{\bmc}
\safemath{\vecd}{\bmd}
\safemath{\vece}{\bme}
\safemath{\vecf}{\bmf}
\safemath{\vecg}{\bmg}
\safemath{\vech}{\bmh}
\safemath{\veci}{\bmi}
\safemath{\vecj}{\bmj}
\safemath{\veck}{\bmk}
\safemath{\vecl}{\bml}
\safemath{\vecm}{\bmm}
\safemath{\vecn}{\bmn}
\safemath{\veco}{\bmo}
\safemath{\vecp}{\bmp}
\safemath{\vecq}{\bmq}
\safemath{\vecr}{\bmr}
\safemath{\vecs}{\bms}
\safemath{\vect}{\bmt}
\safemath{\vecu}{\bmu}
\safemath{\vecv}{\bmv}
\safemath{\vecw}{\bmw}
\safemath{\vecx}{\bmx}
\safemath{\vecy}{\bmy}
\safemath{\vecz}{\bmz}

\safemath{\veczero}{\bmzero}
\safemath{\vecone}{\bmone}
\safemath{\vecxi}{\bmxi}
\safemath{\veclambda}{\bmlambda}
\safemath{\vecmu}{\bmmu}
\safemath{\vectheta}{\bmtheta}
\safemath{\vecphi}{\bmphi}
\safemath{\vecdelta}{\bmdelta}

\safemath{\matA}{\bA}
\safemath{\matB}{\bB}
\safemath{\matC}{\bC}
\safemath{\matD}{\bD}
\safemath{\matE}{\bE}
\safemath{\matF}{\bF}
\safemath{\matG}{\bG}
\safemath{\matH}{\bH}
\safemath{\matI}{\bI}
\safemath{\matJ}{\bJ}
\safemath{\matK}{\bK}
\safemath{\matL}{\bL}
\safemath{\matM}{\bM}
\safemath{\matN}{\bN}
\safemath{\matO}{\bO}
\safemath{\matP}{\bP}
\safemath{\matQ}{\bQ}
\safemath{\matR}{\bR}
\safemath{\matS}{\bS}
\safemath{\matT}{\bT}
\safemath{\matU}{\bU}
\safemath{\matV}{\bV}
\safemath{\matW}{\bW}
\safemath{\matX}{\bX}
\safemath{\matY}{\bY}
\safemath{\matZ}{\bZ}
\safemath{\matzero}{\bmzero}

\safemath{\matDelta}{\bDelta}
\safemath{\matLambda}{\bLambda}
\safemath{\matPhi}{\bPhi}
\safemath{\matSigma}{\bSigma}
\safemath{\matOmega}{\bOmega}
\safemath{\matTheta}{\bTheta}

\safemath{\matidentity}{\matI}
\safemath{\matone}{\matO}

\safemath{\rnda}{A}
\safemath{\rndb}{B}
\safemath{\rndc}{C}
\safemath{\rndd}{D}
\safemath{\rnde}{E}
\safemath{\rndf}{F}
\safemath{\rndg}{G}
\safemath{\rndh}{H}
\safemath{\rndi}{I}
\safemath{\rndj}{J}
\safemath{\rndk}{K}
\safemath{\rndl}{L}
\safemath{\rndm}{M}
\safemath{\rndn}{N}
\safemath{\rndo}{O}
\safemath{\rndp}{P}
\safemath{\rndq}{Q}
\safemath{\rndr}{R}
\safemath{\rnds}{S}
\safemath{\rndt}{T}
\safemath{\rndu}{U}
\safemath{\rndv}{V}
\safemath{\rndw}{W}
\safemath{\rndx}{X}
\safemath{\rndy}{Y}
\safemath{\rndz}{Z}

\safemath{\rveca}{\bimA}
\safemath{\rvecb}{\bimB}
\safemath{\rvecc}{\bimC}
\safemath{\rvecd}{\bimD}
\safemath{\rvece}{\bimE}
\safemath{\rvecf}{\bimF}
\safemath{\rvecg}{\bimG}
\safemath{\rvech}{\bimH}
\safemath{\rveci}{\bimI}
\safemath{\rvecj}{\bimJ}
\safemath{\rveck}{\bimK}
\safemath{\rvecl}{\bimL}
\safemath{\rvecm}{\bimM}
\safemath{\rvecn}{\bimN}
\safemath{\rveco}{\bomO}
\safemath{\rvecp}{\bimP}
\safemath{\rvecq}{\bimQ}
\safemath{\rvecr}{\bimR}
\safemath{\rvecs}{\bimS}
\safemath{\rvect}{\bimT}
\safemath{\rvecu}{\bimU}
\safemath{\rvecv}{\bimV}
\safemath{\rvecw}{\bimW}
\safemath{\rvecx}{\bimX}
\safemath{\rvecy}{\bimY}
\safemath{\rvecz}{\bimZ}

\safemath{\rvecxi}{\bmxi}
\safemath{\rveclambda}{\bmlambda}
\safemath{\rvecmu}{\bmmu}
\safemath{\rvectheta}{\bmtheta}
\safemath{\rvecphi}{\bmphi}

\safemath{\rmatA}{\bimA}
\safemath{\rmatB}{\bimB}
\safemath{\rmatC}{\bimC}
\safemath{\rmatD}{\bimD}
\safemath{\rmatE}{\bimE}
\safemath{\rmatF}{\bimF}
\safemath{\rmatG}{\bimG}
\safemath{\rmatH}{\bimH}
\safemath{\rmatI}{\bimI}
\safemath{\rmatJ}{\bimJ}
\safemath{\rmatK}{\bimK}
\safemath{\rmatL}{\bimL}
\safemath{\rmatM}{\bimM}
\safemath{\rmatN}{\bimN}
\safemath{\rmatO}{\bimO}
\safemath{\rmatP}{\bimP}
\safemath{\rmatQ}{\bimQ}
\safemath{\rmatR}{\bimR}
\safemath{\rmatS}{\bimS}
\safemath{\rmatT}{\bimT}
\safemath{\rmatU}{\bimU}
\safemath{\rmatV}{\bimV}
\safemath{\rmatW}{\bimW}
\safemath{\rmatX}{\bimX}
\safemath{\rmatY}{\bimY}
\safemath{\rmatZ}{\bimZ}

\safemath{\rmatDelta}{\bimDelta}
\safemath{\rmatLambda}{\bimLambda}
\safemath{\rmatPhi}{\bimPhi}
\safemath{\rmatSigma}{\bimSigma}
\safemath{\rmatOmega}{\bimOmega}
\safemath{\rmatTheta}{\bimTheta}

\usepackage{amssymb}
\usepackage{amsfonts}
\usepackage{mathrsfs}
\usepackage{xspace}
\usepackage{bm}
\usepackage{fancyref}
\usepackage{textcomp}

\usepackage{multirow}
\usepackage{stmaryrd}

\newenvironment{textbmatrix}{	\setlength{\arraycolsep}{2.5pt}%
								\big[\begin{matrix}}{\end{matrix}\big]%
								\raisebox{0.08ex}{\vphantom{M}}}

\def\be{\begin{equation}}
\def\ee{\end{equation}}
\def\een{\nonumber \end{equation}}
\def\mat{\begin{bmatrix}}
\def\emat{\end{bmatrix}}
\def\btm{\begin{textbmatrix}}
\def\etm{\end{textbmatrix}}

\def\ba#1\ea{\begin{align}#1\end{align}}
\def\bas#1\eas{\begin{align*}#1\end{align*}}
\def\bs#1\es{\begin{split}#1\end{split}}
\def\bg#1\eg{\begin{gather}#1\end{gather}}
\def\bml#1\eml{\begin{multline}#1\end{multline}}
\def\bi#1\ei{\begin{itemize}#1\end{itemize}}

\newcommand{\lefto}{\mathopen{}\left}

\DeclareMathOperator{\Exop}{\opE}			% expectation operator
\newcommand{\orth}{\perp}					% orthogonal
\newcommand{\Ex}[2]{\ensuremath{\Exop_{#1}\lefto[#2\right]}} 	% expectation
\newcommand{\tp}[1]{\ensuremath{#1^{\text{T}}}} 		% transpose
\newcommand{\herm}[1]{\ensuremath{#1^{\text{H}}}} 	% hermitian transpose
\newcommand{\pinv}[1]{\ensuremath{#1^{\dagger}}} 	% Moore-Penrose pseudo-inverse
\safemath{\dirac}{\delta}					% Dirac delta
\safemath{\krond}{\dirac}					% Kronecker delta
\safemath{\upto}{\uparrow}
\safemath{\downto}{\downarrow}
\safemath{\iu}{j}							% imaginary unit
\safemath{\ev}{\lambda}						% eigenvalue
\safemath{\hilseqspace}{l^{2}}				% Hilbert sequence space
\newcommand{\banachfunspace}[1]{\setL^{#1}}	% Banach function space
\safemath{\hilfunspace}{\banachfunspace{2}}	% Hilbert function space
\safemath{\SNR}{\textit{SNR}} 				% signal to noise ratio
\safemath{\PAR}{\textit{PAR}} 				% signal to noise ratio
\safemath{\No}{N_0}							% noise spectral density
\safemath{\Es}{E_s}							% energy per symbol
\safemath{\Eb}{E_b}							% energy per bit
\safemath{\EbNo}{\frac{\Eb}{\No}}
\safemath{\EsNo}{\frac{\Es}{\No}}

\DeclareMathOperator{\CHop}{\ensuremath{\opH}} % channel operator
\safemath{\tvir}{\rndh_{\CHop}}				% time-varying impulse response
\safemath{\tvtf}{\rndl_{\CHop}}				% 	-''- transfer function
\safemath{\spf}{\rnds_{\CHop}}				% spreading function
\safemath{\bff}{H_{\CHop}}					% bi-freuqency function

\safemath{\ircf}{r_{h}}						% impulse response correlation fn.
\safemath{\tftvcf}{r_{s}}					% scattering function
\safemath{\tfcf}{r_{l}}						% time-frequency correlation fn.
\safemath{\bfcf}{r_{H}}						% bi-frequency correlation fn.

\safemath{\tcorr}{c_h}						% time-correlation function
\safemath{\scf}{c_{s}}						% spreading function
\safemath{\tfcorr}{c_{l}}					% transfer-function correlation
\safemath{\fcorr}{c_{H}}						% frequency-correlation function

\safemath{\mi}{I}							% mutual information
\safemath{\capacity}{C}						% capacity

\safemath{\normal}{\mathcal{N}}			% normal distribution
\safemath{\jpg}{\mathcal{CN}}			% jointly proper Gaussian
\safemath{\mchain}{\leftrightarrow}		% Markov chain
\safemath{\dB}{\,\mathrm{dB}}
\safemath{\dBm}{\,\mathrm{dBm}}
\safemath{\Hz}{\,\mathrm{Hz}}
\safemath{\kHz}{\,\mathrm{kHz}}
\safemath{\MHz}{\,\mathrm{MHz}}
\safemath{\GHz}{\,\mathrm{GHz}}
\safemath{\s}{\,\mathrm{s}}
\safemath{\ms}{\,\mathrm{ms}}
\safemath{\mus}{\,\mathrm{\text{\textmu}s}}
\safemath{\ns}{\,\mathrm{ns}}
\safemath{\ps}{\,\mathrm{ps}}
\safemath{\meter}{\,\mathrm{m}}
\safemath{\mm}{\,\mathrm{mm}}
\safemath{\cm}{\,\mathrm{cm}}
\safemath{\m}{\,\mathrm{m}}
\safemath{\W}{\,\mathrm{W}}
\safemath{\mW}{\, \mathrm{mW}}
\safemath{\J}{\,\mathrm{J}}
\safemath{\K}{\,\mathrm{K}}
\safemath{\bit}{\,\mathrm{bit}}
\safemath{\nat}{\,\mathrm{nat}}

\safemath{\define}{\triangleq}			% definition

\safemath{\equivalent}{\sim}
\safemath{\distas}{\sim}					% distributed according to
\safemath{\sdiff}{\Delta}				% symmetric set difference

\safemath{\reals}{\mathbb{R}}
\safemath{\positivereals}{\reals_{+}}
\safemath{\integers}{\mathbb{Z}}
\safemath{\posint}{\integers_{+}}
\safemath{\naturals}{\mathbb{N}}
\safemath{\posnaturals}{\naturals_{+}}
\safemath{\complexset}{\mathbb{C}}
\safemath{\rationals}{\mathbb{Q}}

\newcommand*{\fancyrefapplabelprefix}{app}		% Appendix
\newcommand*{\fancyrefthmlabelprefix}{thm}		% Theorem
\newcommand*{\fancyreflemlabelprefix}{lem}		% Lemma
\newcommand*{\fancyrefcorlabelprefix}{cor}		% Corollary
\newcommand*{\fancyrefdeflabelprefix}{def}		% Definition
\newcommand*{\fancyrefproplabelprefix}{prop}		% Proposition
\newcommand*{\fancyrefexmpllabelprefix}{exmpl}
\newcommand*{\fancyrefalglabelprefix}{alg}		% Algorithm
\newcommand*{\fancyreftbllabelprefix}{tbl}		% Algorithm

\frefformat{vario}{\fancyrefseclabelprefix}{Sec.~#1}
\frefformat{vario}{\fancyrefthmlabelprefix}{Thm.~#1}
\frefformat{vario}{\fancyreftbllabelprefix}{Table~#1}
\frefformat{vario}{\fancyreflemlabelprefix}{Lemma~#1}
\frefformat{vario}{\fancyrefcorlabelprefix}{Corollary~#1}
\frefformat{vario}{\fancyrefdeflabelprefix}{Def.~#1}
\frefformat{vario}{\fancyreffiglabelprefix}{Fig.~#1}
\frefformat{vario}{\fancyrefapplabelprefix}{Appendix~#1}
\frefformat{vario}{\fancyrefeqlabelprefix}{(#1)}
\frefformat{vario}{\fancyrefproplabelprefix}{Prop.~#1}
\frefformat{vario}{\fancyrefexmpllabelprefix}{Example~#1}
\frefformat{vario}{\fancyrefalglabelprefix}{Alg.~#1}

 \newtheorem{thm}{Theorem}
 \newtheorem{prop}{Proposition}

 \newtheorem*{remark*}{Remark}
\safemath{\dictab}{[\,\dicta\,\,\dictb\,]}

\safemath{\ysig}{\bmy}
\safemath{\ysighat}{\hat{\ysig}}
\safemath{\ysigdim}{M}
\safemath{\xsig}{\bmx}
\safemath{\xsigdim}{N}
\safemath{\nx}{n_x}
\safemath{\zsig}{\bmz}
\safemath{\zsigdim}{\ysigdim}
\safemath{\rsig}{\bmr}
\safemath{\Adict}{\bA}
\safemath{\Adicttilde}{\widetilde{\Adict}}
\safemath{\Adictdim}{\outputdim\times\xsigdim}
\safemath{\avec}{\bma}
\safemath{\avectilde}{\tilde{\avec}}
\safemath{\Bdict}{\bB}
\safemath{\Bdicttilde}{\widetilde{\Bdict}}
\safemath{\Cdict}{\bC}
\safemath{\cvec}{\bmc}
\safemath{\Ddict}{\bD}
\safemath{\Ddictdim}{\ysigdim\times\xsigdim}
\safemath{\dvec}{\bmd}
\safemath{\Ddicttilde}{\widetilde{\bD}}
\safemath{\Bonb}{\bB}
\safemath{\bvec}{\bmb}
\safemath{\Bonbdim}{\ysigdim\times\ysigdim}
\safemath{\noise}{\bmn}
\safemath{\noisedim}{\ysigim}
\safemath{\err}{\bme}
\safemath{\errdim}{\ysigdim}
\safemath{\errset}{\setE}
\safemath{\nerr}{n_e}
\safemath{\delop}{\bP_\errset}
\safemath{\delopc}{\bP_{{\errset}^c}}

\safemath{\cplxi}{\imath}
\safemath{\cplxj}{\jmath}
\safemath{\dict}{\matD}
\safemath{\inputdim}{N}		% number of columns of dictionary D
\safemath{\outputdim}{M}		%number of rows of dictionary D
\safemath{\sparsity}{S}	%sparsity
\safemath{\inputdimA}{{N_a}}	%total number of elements in dictionary A
\safemath{\inputdimB}{{N_b}}	%total number of elements in dictionary B
\safemath{\elemA}{{n_a}}	%number of elements chosen from dictionary A
\safemath{\elemB}{{n_b}}	%number of elements chosen from dictionary B
\safemath{\resA}{\matR_a}	%restriction map to elements of dictionary A
\safemath{\resB}{\matR_b}	%restriction map to elements of dictionary B
\safemath{\subD}{\matS} %subdictionary
\safemath{\subA}{\matS_a} %subdictionary part of A
\safemath{\subB}{\matS_b} %subdictionary part of B
\safemath{\dicta}{\matA} 	% first subdictionary
\safemath{\dictb}{\matB} 	% second subdictionary
\safemath{\hollowS}{H}
\safemath{\hollowA}{H_a}
\safemath{\hollowB}{H_b}
\safemath{\cross}{Z}
\safemath{\coh}{\mu_d}			% coherence dictionary
\safemath{\coha}{\mu_a}			% coherence first subdictionary
\safemath{\cohb}{\mu_b}			% coherence second subdictionary
\safemath{\mubs}{\nu}	%block sub-coherence
\safemath{\cohm}{\mu_m} %mutual coherence
\safemath{\dictset}{\setD}	% set of dictionaries
\safemath{\dictsetp}{\dictset(\coh,\coha,\cohb)}	% set of dictionaries parametrized
\safemath{\dictsetgen}{\dictset_\text{gen}}
\safemath{\dictsetgenp}{\dictsetgen(\coh)}
\safemath{\dictsetonb}{\dictset_\text{onb}}
\safemath{\dictsetonbp}{\dictsetonb(\coh)}

\safemath{\leftside}{U}
\safemath{\rightsideA}{R_a}
\safemath{\rightsideB}{R_b}

\safemath{\indexS}{\setI_S} %set of indices participating in sub-dictionary S

\safemath{\na}{n_a}			% cardinality of set of linearly independent columns of first dictionary
\safemath{\nb}{n_b}			% cardinality of set of linearly independent columns of second dictionary
\safemath{\coeffa}{p_i}	%coefficients for columns of A
\safemath{\coeffb}{q_j}	%coefficients for columns of B
\safemath{\seta}{\setP}		% set of linearly independent columns of A
\safemath{\setb}{\setQ}     % set of linearly independent columns of B
\safemath{\setw}{\setW}	%set of n largest elements of w
\safemath{\setz}{\setZ}	%set of L-n largest elements of z
\safemath{\cola}{\veca}		% generic element of the dictionary A
\safemath{\colb}{\vecb}		% generic element of the dictionary B
\safemath{\cold}{\vecd}		% generic element of the dictionary D
\safemath{\inputvec}{\vecx} 	%coefficient vector (input)
\safemath{\error}{\vece}	%error vector
\safemath{\noiseout}{\vecz} 	%noisy output vector
\safemath{\inputvecel}{x}
\safemath{\inputveca}{\vecx_a}
\safemath{\inputvecb}{\vecx_b}
\safemath{\outputvec}{\vecy}	%output of Dictionary
\safemath{\lambdamin}{\lambda_{\mathrm{min}}}
\safemath{\elltwo}{\ell_2}
\safemath{\ellone}{\ell_1}
\safemath{\ellzero}{\ell_0}
\safemath{\ellinf}{\ell_\infty}
\safemath{\ellinftilde}{\ell_{\widetilde\infty}}
\safemath{\licard}{Z(\coh,\coha,\cohb)}
\safemath{\xsol}{\hat{x}}
\safemath{\xbord}{x_b}		%Solution at the border
\safemath{\xstat}{x_s}		%Solution stationary in l0 prob
\safemath{\xstatLone}{\tilde{x}_s}
\safemath{\order}{\mathcal{O}} %order notation (big O)
\safemath{\scales}{\Theta} %scales as
\safemath{\ones}{\mathbf{1}} %all ones matrix
\safemath{\zeroes}{\mathbf{0}} %all zeroes matrix
\safemath{\thlone}{\kappa(\coh,\cohb)} %treshold l1 problem
\safemath{\constoneA}{\delta} %constant in l1 theorem to save space
\safemath{\constoneB}{\epsilon} %constant in l1 theorem to save space
\safemath{\nlarge}{L}				   %num large elements
\safemath{\sumlarge}{S_\nlarge}
\safemath{\maxlarger}{P_\nlarge}	   % maximum in Gribonval and Nielsen
\safemath{\Pzero}{\textrm{P0}}	
\safemath{\Pone}{\textrm{P1}}
\safemath{\vecfir}{\vecw}			 % \vecv element of the kernel of the dictionary, \vecv=[\vecfir \vecsec]
\safemath{\vecsec}{\vecz}
\safemath{\elvecfir}{w}              % element of vecfir
\safemath{\elvecsec}{z}				 % element of vecsec
\safemath{\nlargefir}{n}
\safemath{\normout}{\gamma}
\safemath{\auxfun}{h}
\safemath{\supp}{\textrm{supp}}%support

\safemath{\indexa}{\ell}
\safemath{\indexb}{r}
\safemath{\indexc}{i}
\safemath{\indexd}{j}

\safemath{\project}{P}%projector

\begin{document}
\title{Active Eavesdropper Mitigation\\via Orthogonal Channel Estimation}

\author{%
  \IEEEauthorblockN{Gian Marti and Christoph Studer}
  \IEEEauthorblockA{\em Department of Information Technology and Electrical Engineering, ETH Zurich, Switzerland\\
              email: gimarti@ethz.ch and studer@ethz.ch}
}

\maketitle

\begin{abstract}
Beamforming is a powerful tool for physical layer security, as it 
can be used for steering
signals towards legitimate receivers and away from eavesdroppers. 
An active eavesdropper, however, can interfere with the pilot phase that the transmitter needs to 
acquire the channel knowledge necessary for beamforming. By doing so, the eavesdropper can 
make the transmitter form beams towards the eavesdropper rather than towards the legitimate receiver.
To mitigate active eavesdroppers, we propose VILLAIN, a novel channel estimator that uses secret pilots. 
When an eavesdropper interferes with the pilot phase, VILLAIN produces a channel estimate that
is orthogonal to the eavesdropper's channel (in the noiseless case). We prove that beamforming based on this channel estimate delivers the highest 
possible signal power to the legitimate receiver without delivering any signal power to the eavesdropper. 
Simulations show that VILLAIN mitigates active eavesdroppers also in the noisy~case. 

\end{abstract}

\vspace{-.5mm}
\section{Introduction}\label{sec:intro}

Security is a concern of paramount importance in modern communication systems \cite{whitman2021principles}. 
Physical layer security (PLS) is emerging as a powerful alternative to classical cryptography \cite{bloch2011physical, wu2018survey}. 
While classical cryptography is based on the assumption that certain computational 
problems are hard, PLS builds on the characteristics of the channel itself and offers information-theoretic security. 
In multi-antenna transmission systems, PLS can leverage beamforming to steer communication signals 
towards the intended recipient and away from an eavesdropper~\cite{li2007secret}.

Beamforming requires the transmitter to know the channels to the receivers---to 
the legitimate receiver for steering signals towards it and to 
the eavesdropper for steering signals away from it. 
Most research on eavesdropper mitigation simply assumes perfect knowledge of all channels \cite{shafiee2007achievable, li2007secret, rezki2011finite, reboredo2013filter}.
However, it is unclear how the transmitter could obtain channel knowledge
of an eavesdropper, which---if not declared otherwise---is usually understood to be 
a \emph{passive} eavesdropper, i.e., one which emits no signals. 
Thankfully, transmitters with a large number of antennas (as 
are used, e.g., in the massive multiple-input multiple-output (MIMO) downlink) 
are intrinsically resistant to passive eavesdropping due to the narrow 
beams that such transmitters form towards the receiver. 
Such narrow beams entail significantly higher signal strength at the legitimate 
receiver than at the eavesdropper \cite{kapetanovic2015physical, bereyhi2018robustness, bereyhi2019robustness}.

In contrast to passive eavesdroppers, 
\emph{active} eavesdroppers try to influence communication in their favor 
by emitting signals themselves.
For instance, an active eavesdropper may contaminate the channel estimation (or \emph{pilot})
phase such that the transmitter 
forms beams towards the eavesdropper instead of the receiver 
\cite{zhou2012pilot, wu2016secure, bereyhi2020secure}. 
This pilot contamination renders active eavesdroppers effective also against transmitters with 
many antennas. 
On the flip side, the signals that an active eavesdropper emits
give the transmitter an opportunity to \emph{somehow} estimate the eavesdropper's 
channel and then using beamforming to steer signals away. 
Much research on active eavesdropper mitigation therefore simply assumes
that the channel of an active eavesdropper is perfectly 
\cite{li2017mimo, cho2020zero, si2020cooperative, cho2021cooperative, zhou2022caching} 
or imperfectly \cite{yu2020robust, jia2023star} known at the transmitter,  
or that at least its second-order statistics are known \cite{wu2016secure, bereyhi2020secure}.
\emph{How} this channel knowledge should be obtained is usually not
discussed, however. In particular, a sophisticated active eavesdropper might only transmit
during the pilot phase and thus never provide the transmitter with snapshots
of its channel that are uncontaminated by the pilot 
signals of the legitimate receiver.

\vspace{-.5mm}
\subsection{Contributions}
We propose VILLAIN (short for eaVesdropper resILient channeL estimatIoN)
for active eavesdropper mitigation in the single-user MIMO downlink. 
When an active eavesdropper contaminates the pilot phase, VILLAIN produces an estimate of the legitimate
receiver's channel that is orthogonal to the eavesdropper's channel (if the noise at the basestation (BS) is negligible). 
We prove that using this channel estimate for maximum-ratio transmission (MRT) results in a beamformer that is optimal
in the sense that it delivers the highest possible signal power to the legitimate
receiver while simultaneously ensuring that the received signal power at the eavesdropper is zero.
VILLAIN therefore guarantees perfect secrecy against active eavesdroppers,\footnote{By perfect secrecy, 
we mean that $H(s|y_{\text{ed}})=H(s)$, where $s$ and $y_{\text{ed}}$ are defined in \fref{sec:model}, 
and where $H(\cdot)$ and $H(\cdot|\cdot)$ denote (conditional) entropy.} 
and it does so without requiring wiretap coding. 
Using numerical simulations, we show that VILLAIN succeeds in mitigating active 
eavesdroppers also when the noise at the BS is not negligible. 

\vspace{-.5mm}
\subsection{Notation}
Column vectors and matrices are denoted by lowercase~boldface (e.g.,~$\bma$) and uppercase boldface (e.g. $\bA$) letters, respectively.
The transpose is denoted $\tp{(\cdot)}$, the complex conjugate~$(\cdot)^\ast$, the conjugate transpose $\herm{(\cdot)}$, 
and the Moore-Penrose pseudoinverse~$\pinv{(\cdot)}$.
The Frobenius-norm is $\|\cdot\|_F$, the $2$-norm $\|\cdot\|_2$, 
and the absolute value~$|\cdot|$.
The subspace spanned~by~$\bma$ is $\text{span}(\bma)$ and its orthogonal complement is $\text{span}(\bma)^\orth$.
The circularly-symmetric complex Gaussian distribution with variance $Q$ is $\setC\setN(0,Q)$.
The expectation operator is~$\Ex{}{\cdot}$.

\newpage
\section{System Model}\label{sec:model}

We consider the case where a $B$-antenna BS wants to transmit data (from a constellation $\setS$ 
with unit average symbol energy) 
to a single-antenna user equipment (UE) in the presence of a single-antenna eavesdropper.
The receive signals at the UE and the eavesdropper can be written as
\begin{align}
y_{\text{ue}} &= \tp{\bmh}\bmx + n_{\text{ue}}, \\
y_{\text{ed}} &= \tp{\bmj}\bmx + n_{\text{ed}},
\end{align}
respectively. Here, $\bmx\in\opC^B$ is the BS transmit vector that must satisfy a power constraint 
$\Ex{}{\|\bmx\|_2^2}\leq P$, $\tp{\bmh},\tp{\bmj}\in\opC^B$ are the downlink channel vectors (which include 
the effects of large-scale as well as of small-scale fading) between
the BS and the UE and the eavesdropper, respectively, and $n_{\text{ue}}\sim\setC\setN(0,\textsf{N}_{\text{ue}})$
and $n_{\text{ed}}\sim\setC\setN(0,\textsf{N}_{\text{ed}})$ model the noise at the UE and the eavesdropper, 
respectively. 
The transmit vector $\bmx$ is a linear function of the data symbol $s\in\setS$ to be sent to the UE, i.e., 
\begin{align}
	\bmx = \bmw s, 
\end{align}
where $\bmw$ is the BS's precoding vector.\footnote{
Since we assume that the constellation $\setS$ has unit average symbol energy, 
the power constraint $\Ex{}{\|\bmx\|_2^2}\leq P$ is equivalent to $\|\bmw\|_2^2 \leq P$.
} 
This precoding vector has two objectives: 
First, it should ensure that the UE can easily recover $s$ based on $y_{\text{ue}}$. 
Second, it should ensure that the eavesdropper \emph{cannot} recover $s$ based on $y_{\text{ed}}$ 
(not even if the eavesdropper knows both $\bmj$ and $\bmw$). 

The BS determines its precoding vector $\bmw$ based on a pilot phase in which the UE transmits
a length-$T$ pilot sequence $\bms_T\in\opC^T$ that is known to the BS. In other words, 
$\bmw = f(\bY_T)$
for some function $f: \opC^{B\times T} \to \opC^B$, where $\bY_T\in\opC^{B\times T}$ is the BS's pilot receive signal. 
We assume that $\bms_T$ is \emph{secret} 
(i.e., unknown to the eavesdropper) and potentially \emph{random}.\footnote{
To prevent the eavesdropper from learning the pilot sequence, 
$\bms_T$ should be changed every coherence time. For information-theoretic
security, this would require reading from a one-time pad (OTP). 
However, this OTP could be much shorter than a message-encrypting OTP, 
since pilots are transmitted~infrequently.
Moreover, even if one abandons information-theoretic security and uses a
cryptographic random number generator (CRNG) to update $\bms_T$, 
the resulting security is much better than in classical cryptography, 
since the eavesdropper has to break the CRNG \emph{in real time}
to create an attack opening (while in classical cryptography, 
the received message can be stored and cracked offline).
}

In a no-eavesdropper or passive eavesdropper scenario, the receive matrix $\bY_T$ from the pilot phase can be written as
\begin{align}
	\bY_T = \bmh \tp{\bms_T} + \bN_T, \label{eq:pilot_no_ed}
\end{align}
where the UE uplink channel vector $\bmh$ is the transpose of the UE downlink channel vector due to channel reciprocity, 
and where $\bN_T\stackrel{\text{i.i.d.}}{\sim}\setC\setN(0,\textsf{N}_{\text{bs}})$ models the receive noise at the BS. 

The precoding vector $\bmw$ is often determined by first forming an estimate $\hat\bmh=g(\bY_T)$ of the 
UE channel vector $\bmh$, and then setting $\bmw=h(\hat\bmh)$ for some function $h$ (i.e., $f=h\circ g$).
A classic example is least squares (LS) channel estimation
\begin{align}
	\hat\bmh = g(\bY_T) =  \bY_T \pinv{(\tp{\bms_T})} \label{eq:simo_ls}
\end{align}
followed by maximum ratio transmission (MRT) precoding
\begin{align}
	\bmw = h(\hat\bmh) = \sqrt{P}\,\hat\bmh^{\ast}/\|\hat\bmh\|_2. \label{eq:mrt}
\end{align}
The UE can then simply estimate $s$ by rescaling $y_{\text{ue}}$ as 
\begin{align}
	\hat{s} = \beta y_{\text{ue}},
\end{align}
where $\beta=1/|\tp{\bmh}\bmw|$ recovers the scale of the transmit signal.\footnote{This 
way of expressing the scaling factor $\beta$ assumes---unrealistically---that the UE knows both $\bmh$ and $\bmw$. 
However, $\beta$ depends primarily on the large-scale fading between the UE and the BS, which changes slowly in 
time. We therefore assume that the UE can estimate $\beta$ from the receive signals \cite{jacobsson16d}.}
With these choices of $\bmw$ and $\beta$, we have $\hat{s}\to s$ as $\|\bN_T\|_F \to 0$.

In this paper, we consider an \emph{active} eavesdropper that transmits a signal $\bmz\in\opC^T$ during
the pilot phase to make the BS use a precoding vector $\bmw$ that makes it easy 
for the eavesdropper to detect $s$ based on $y_{\text{ed}}$.
Thus, the pilot receive signal does not have the form of \eqref{eq:pilot_no_ed}, 
but instead can be written~as
\begin{align}
	\bY_T = \bmh \tp{\bms_T} + \bmj \tp{\bmz} + \bN_T. \label{eq:pilot_io}	
\end{align}
In the active eavesdropper literature, it is typically assumed that the eavesdropper knows 
the pilot sequence $\bms_T$. In that case, it is natural for the eavesdropper to also transmit
the pilot sequence (potentially at higher power), i.e., $\tp{\bmz}=\alpha\tp{\bms_T}$ 
for some $\alpha\geq1$. A BS that uses an LS channel estimator as in \eqref{eq:simo_ls} 
with MRT beamforming as in \eqref{eq:mrt} will then effectively form a least-square estimate not of $\bmh$, 
but of $\bmh + \alpha\bmj$. As $\alpha\to\infty$, this becomes a (scaled) estimate of $\bmj$, 
so that the BS optimizes its precoding vector for the eavesdropper instead of the UE 
and, consequently, forms its beam towards the eavesdropper rather than towards the UE.

In contrast, we assume the eavesdropper does not know $\bms_T$, so that $\bmz$ cannot depend on $\bms_T$. 
However, the eavesdropper can still influence the BS to its advantage 
by sending $\bmz\stackrel{\text{i.i.d.}}{\sim}\setC\setN(0,Q)$ for some $Q>0$.
In that case, the LS channel estimator in~\eqref{eq:simo_ls} gives 
$\bmh\tp{\bms_T}\pinv{(\tp{\bms_T})} = \bmh$ in the UE term, 
but $\bmj\tp{\bmz}\pinv{(\tp{\bms_T})}= \omega \bmj$ in the eavesdropper term, 
where $\omega \sim \setC\setN(0,Q/\|\bms_T\|_2^2)$.
So, even if the eavesdropper transmits Gaussian noise, 
a BS with a LS channel estimator effectively forms a least-square estimate of $\bmh+\omega\bmj$, 
where $\omega$ is complex Gaussian with variance $Q/\|\bms_T\|_2^2$.
If $Q\gg \|\bms_T\|_2^2$, then with high probability $|\omega|>1$, 
so that the BS chooses $\bmw$ mainly as a function of the eavesdropper channel vector $\bmj$ and not the UE channel vector~$\bmh$.

\section{VILLAIN: A Channel Estimator\\for Active Eavesdropper Mitigation}
Before we present VILLAIN, a remark is in order. 
VILLAIN is a channel estimator---how can a channel estimator mitigate an eavesdropper? 
The answer is that VILLAIN does not produce an unbiased estimate 
of $\bmh$, but one that is projected onto (an estimate of) the orthogonal 
complement $\text{span}(\bmj)^\orth$ of the eavesdropper subspace $\text{span}(\bmj)$. 
The following result shows that if such a channel estimate is combined 
with MRT precoding as in \eqref{eq:mrt}, then the eavesdropper receives no signal.
\begin{prop} \label{prop:simo_ortho_guarantee}
	If the BS obtains a channel estimate $\hat\bmh\neq\mathbf{0}$
	from the image of the projection $\bP=\bI-\bmj\pinv{\bmj}$ onto
	$\text{span}(\bmj)^\orth$,  (i.e., $\hat\bmh=\bP\tilde\bmh$ 
	for some $\tilde\bmh\in\opC^B$) and uses MRT precoding as in \eqref{eq:mrt},
	then the eavesdropper receives no signal, $\bmj^{\textnormal{T}}\bmw = 0$.
\end{prop}
All proofs are in the Appendix.
Motivated by this result, we now present VILLAIN.
In VILLAIN, the UE sends a \emph{redundant} pilot sequence, i.e., 
a pilot sequence of length $T>1$. 
As we will see, this redundancy allows the BS to estimate two things: 
the eavesdropper subspace and the projection of the UE's channel onto the orthogonal complement of the eavesdropper~subspace. 

Let the pilot phase be given as in \eqref{eq:pilot_io}. 
Then VILLAIN solves
\begin{align}
	\min_{\substack{\tilde\bmh \,\in\, \opC^{B},\\ \tilde\bP \,\in\, \mathscr{G}_{B-1}(\opC^B)}} 
	\big\| \tilde\bP\bY_T - \tilde\bmh \tp{\bms_T} \big\|_F^2. \label{eq:simo_pilot_problem}
\end{align}
Here, $\mathscr{G}_{B-1}(\opC^B) = \{\bI_B - \bma\pinv{\bma} : \bma\in\opC^B\}$ is the Grassmannian manifold, 
i.e., the set of orthogonal projections onto $(B-1)$-dimensional subspaces of $\opC^B$. 
The channel estimate $\hat\bmh$ that is obtained from solving \eqref{eq:simo_pilot_problem} can then be used 
for MRT precoding as in \eqref{eq:mrt}.
Even though the problem in \eqref{eq:simo_pilot_problem} is non-convex, 
it has a closed-form solution:
\begin{prop} \label{prop:solution_simo_pilot_problem}
The problem in \eqref{eq:simo_pilot_problem} is solved by 
\begin{align}
	\hat\bP = \bI_B - \bmu \bmu^{\textnormal{H}} \quad\text{and}\quad \hat\bmh = \hat\bP\bY_T\pinv{({\bms}^{\textnormal{T}}_T)}, \label{eq:solution_simo_pilot_problem}
\end{align}
where $\bmu\in \opC^B$ is the left singular vector which corresponds to the 
largest singular value of $\bY_T(\bI_T-\pinv{({\bms}^{\textnormal{T}}_T)}{\bms}^{\textnormal{T}}_T)$.
\end{prop}

In \eqref{eq:solution_simo_pilot_problem}, $\bmu$ should be~understood 
as an estimate of the eavesdropper subspace 
(i.e., $\bmu \approx \alpha\bmj$ for some $\alpha\in\opC$), 
$\hat\bP$ is the orthogonal projection onto the orthogonal complement of 
that subspace (i.e., $\hat\bP\approx\bI_B-\bmj\pinv{\bmj}$), 
and $\hat\bmh$ is an estimate of the projection of the
UE's channel vector onto the orthogonal complement of the eavesdropper 
subspace (i.e., $\hat\bmh\approx(\bI_B-\bmj\pinv{\bmj})\bmh$). 
If the pilot sequence $\bms_T$ is chosen at random and unknown to the eavesdropper, 
and if there is no noise at the BS, then we have the following~guarantee: 

\begin{thm} \label{thm:best_gain}
	Assume that $T\!>\!1$, that $\bms_T\!\stackrel{\textnormal{i.i.d.}}{\sim}\!\setC\setN(0,1)$ 
	and \mbox{$\bmz\!\neq\!\mathbf{0}$} are independent of each other, and	that $\mathsf{N}_{\textnormal{bs}}=0$. 
	Then, almost surely, the precoding vector $\bmw$ that results from 
	the VILLAIN channel estimator in conjunction with MRT precoding solves
	\begin{align}
		\max_{\tilde\bmw\in\opC^{B}:\|\tilde\bmw\|_2^2\leq P} 
		|\bmh^{\textnormal{T}}\tilde\bmw|^2 \quad \text{such that} \quad \bmj^{\textnormal{T}}\tilde\bmw=0.
		\label{eq:opt_problem_ue_rec_power}
	\end{align}
	That is, $\bmw$ achieves the highest UE receive signal power of all vectors that achieve zero eavesdropper receive signal~power.
	The delivered signal power is $|\bmh^{\textnormal{T}}\bmw|^2 = \|(\bI_B-\bmj\pinv{\bmj})\bmh\|_2^2 P$.
\end{thm}

Note that the mitigation strategy pursued here---maximizing the receive power at the receiver while ensuring that the eavesdropper
receives \emph{no} signal---is suboptimal in the sense that it generally does not achieve the 
secrecy capacity~\cite{khisti2010secure}. However, VILLAIN has the following advantages: 
First, under the conditions of \fref{thm:best_gain}, it achieves perfect secrecy \emph{without} requiring a priori 
information of the UE's or the eavesdropper's channel. 
Second, it then achieves perfect secrecy using off-the-shelf MRT precoding, 
without requiring wiretap coding. The only price to be paid is a randomized pilot phase of length at least two, 
and computing the VILLAIN channel estimate in \fref{prop:solution_simo_pilot_problem}. 
Its computational complexity is dominated by the SVD, whose complexity scales with 
$\max\{B,T\}\min\{B,T\}^2$.

\section{Simulation Results}

\subsection{Line-of-Sight Channel Without Noise at the BS}

For illustrative purposes, we start by considering
a textbook line-of-sight (LoS) channel, where a BS with $B=8$ antennas 
arranged as a uniform linear array (ULA) with antennas spaced at half a wavelength
is located at the coordinate origin, while the UE and the eavesdropper are located 
in far-field at degrees of $\theta_{\text{UE}}=70^\circ$ and
$\theta_{\text{ED}}=20^\circ$, respectively.
We assume that there is no noise at the BS ($\mathsf{N}_{\text{bs}}=0$),
that the BS power constraint is $P=1$, and that the channel gains of the UE and the eavesdropper 
are $\|\bmh\|_2 = \|\bmj\|_2 = 1$. 
A textbook LoS channel vector $\bmg$ with unit gain can be written in dependence of the angle $\phi$ 
(relative the the ULA)~as
\begin{align}
	\bmg(\phi) &= \frac{1}{\sqrt{B}}\tp{\Big[1, e^{-i\pi\cos(\phi)}, \dots, e^{-i\pi\cos(\phi)(B-1)} \Big]},
\end{align}
and we have $\bmh=\bmg(\theta_{\text{UE}})$ and $\bmj=\bmg(\theta_{\text{ED}})$.

\begin{figure*}[tp]
\centering
\subfigure[Passive eavesdropper and LS channel est.]{
\includegraphics[width=0.31\textwidth]{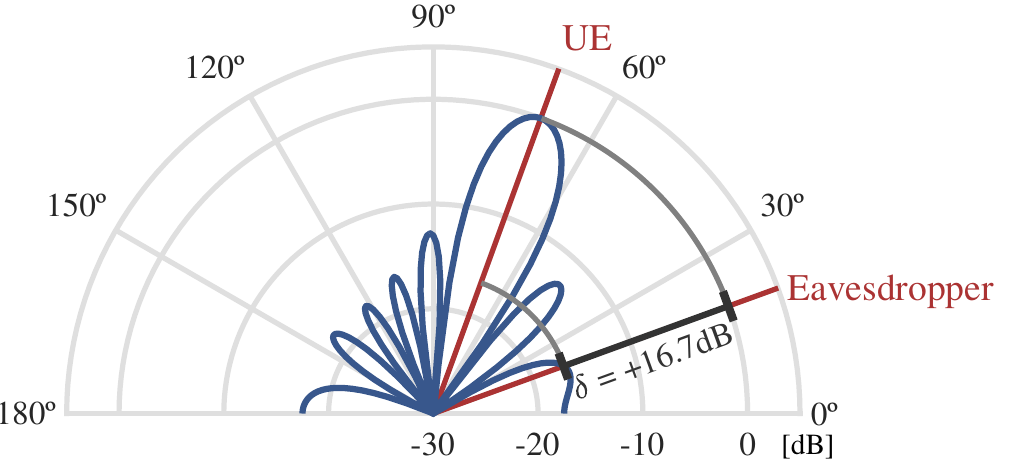}
\label{fig:example:passive}
}
\subfigure[Active eavesdropper and LS channel est.]{
\includegraphics[width=0.31\textwidth]{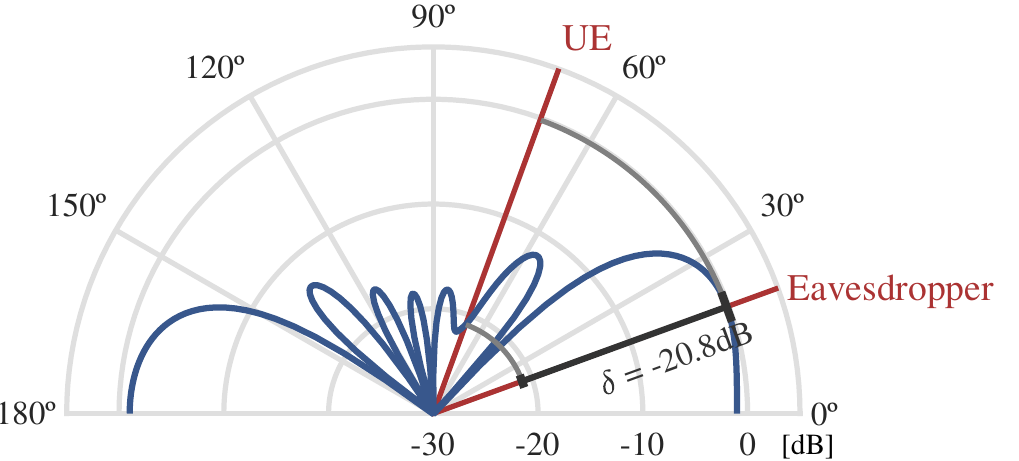}
\label{fig:example:active}
}
\subfigure[Active eavesdropper and VILLAIN channel est.]{
\includegraphics[width=0.31\textwidth]{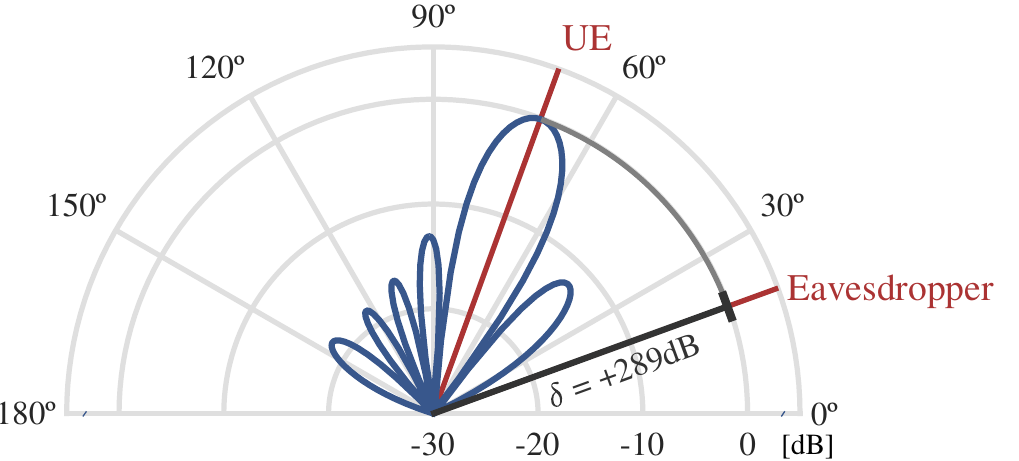}
\label{fig:example:act_villain}
}
\caption{
Signal receive power (in dB) as a function of the incidence angle $\phi$ for three different scenarios.
The BS has $B=8$ antennas arranged as a ULA. We assume a textbook LoS channel with the UE and the eavesdropper 
being in far-field at $70^\circ$ and $20^\circ$, respectively. 
The performance of a precoding can be characterized in terms of the ratio 
$\delta$ between the signal receive power at the UE and at the eavesdropper. 
VILLAIN effectively mitigates the active eavesdropper.
}
\label{fig:example_det}
\vspace{-3mm}
\end{figure*}

We consider three different scenarios in which the BS computes its precoding 
vector $\bmw$. For each of the scenarios,~\fref{fig:example_det} shows the receive power 
$\Ex{}{|\tp{\bmg}(\phi)\bmx|^2} = |\tp{\bmg}(\phi)\bmw|^2$ (in dB) as a function of $\phi$.
For each of the scenarios, we also compute the \emph{advantage} $\delta$, which we define as the 
ratio between the power received at the UE and the power received at the eavesdropper, 
\begin{align}
	\delta \triangleq \frac{|\tp{\bmh}\bmw|^2}{|\tp{\bmj}\bmw|^2}. \label{eq:advantage}
\end{align}
If the noise at the UE and the eavesdropper are equally strong, $\mathsf{N}_\text{ue}=\mathsf{N}_\text{ed}$, 
then $\delta>1$ implies that the secrecy capacity \emph{for that precoding vector} $\bmw$ 
is positive; $\delta\leq1$ implies that it is zero.
The three considered scenarios are as follows:
\subsubsection{Passive Eavesdropper and LS channel estimation}
In this scenario, the eavesdropper does not transmit during the pilot phase 
in which the UE sends a pilot sequence of length $T=8$. The BS uses the LS channel estimator of \eqref{eq:simo_ls}
with the MRT precoder of \eqref{eq:mrt}.
The signal receive strength (as a function of~$\phi$) for the resulting precoder is shown in 
\fref{fig:example:passive}. The receive strength is highest at $\phi=\theta_{\text{UE}}$, where a
gain of $0$\,dB is achieved. In contrast, the receive strength at the eavesdropper is significantly lower, 
and an advantage of $\delta=+16.7$\,dB is achieved. This confirms the claim of 
\cite{kapetanovic2015physical, bereyhi2018robustness, bereyhi2019robustness} 
that multi-antenna precoding protects naturally against passive eavesdroppers. 
\subsubsection{Active Eavesdropper}
This scenario differs from the previous one in that the eavesdropper transmits i.i.d. circularly-symmetric 
Gaussian samples (independent of~$\bms_T$, so that the eavesdropper need not know $\bms_T$)
during the pilot phase, 
at $25$\,dB higher expected power than $\bms_T$. 
The BS uses the same LS channel estimator of \eqref{eq:simo_ls} and  MRT-precoder of \eqref{eq:mrt} as in the first scenario.
The signal receive strength for the resulting precoding is shown in \fref{fig:example:active}.
Since the pilot receive signal is dominated by the eavesdropper, the signal receive strength is
highest in the direction of the eavesdropper, where a gain of almost $0$\,dB is achieved. 
In contrast, the signal receive strength at the UE is much lower, resulting in a negative gain of 
$\delta=-20.8$\,dB. This result shows a clear need for active eavesdropper mitigation
if physical-layer security is desired.

\subsubsection{Active Eavesdropper with VILLAIN}
The third scenario is identical to the second one, except that the BS now uses the VILLAIN 
channel estimator (together with the MRT precoder of \eqref{eq:mrt}).
The signal receive strength of the resulting precoding vector is plotted in 
\fref{fig:example:act_villain}.
The results show that VILLAIN succeeds in mitigating the active eavesdropper:
Even though the pilot receive signal is dominated by the eavesdropper's contribution, 
the signal receive strength is highest at the UE, where a gain of almost $0$\,dB is achieved. 
In contrast, the receive strength at the eavesdropper is now so low that it does not even 
show on the axis, and an advantage of $\delta=+289$\,dB is achieved. 
In theory, the advantage in such a noiseless scenario should be $\delta=+\infty$\,dB, but the floating-point
accuracy of MATLAB simulations limits the advantage to a finite value.

\subsection{QuaDRiGa UMa Channels With Noise at the BS}
\begin{figure}
	\centering
	\includegraphics[width=0.75\columnwidth]{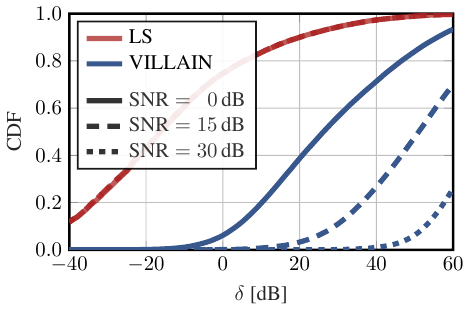}
	\vspace{-2mm}
	\caption{CDF of VILLAIN and LS channel estimation for different noise levels at the BS. The eavesdropper's transmit signal is 30dB stronger
	than the UE's.}
	\vspace{-2mm}
	\label{fig:cdf}
\end{figure}

We now simulate a more realistic scenario that considers noise at the BS 
and uses channel vectors that are generated using QuaDRiGa \cite{jaeckel2014quadriga} with a 3GPP 38.901 urban macrocellular (UMa) channel model \cite{3gpp22a}.
The carrier frequency is $2$\,GHz, the BS has $B=16$ antennas arranged
as a ULA spaced at half a wavelength, 
and the UE and the eavesdropper are uniformly and independently placed 
at a distance between $10$\,m and $100$\,m in a $120^{\circ}$ sector in front of the BS.
We compare the performance of VILLAIN to the performance of a conventional LS channel estimator. 
In both cases, the UE transmits an i.i.d. $\setC\setN(0,\Es)$ pilot sequence of length $T=4$. 
The eavesdropper transmits i.i.d. $\setC\setN(0,Q)$ samples 
at $30$\,dB higher transmit power than the UE during the pilot phase
(i.e., $10\log_{10}(Q/\Es)=30$\,dB). 
We quantify the BS noise variance $\mathsf{N}_\text{bs}$ relative to the \emph{transmit} signal power~$\Es$,
where we define the signal-to-noise ratio as $\textit{SNR}=\Es/\mathsf{N}_\text{bs}$.
We consider three different noise levels: $\textit{SNR}=0$\,dB, $\textit{SNR}=15$\,dB, and $\textit{SNR}=30$\,dB.
\fref{fig:cdf} shows the cumulative distribution function (CDF) for the different channel estimators and 
the different SNRs.\footnote{
These CDFs take into account the large channel gain variation between the UE 
and the eavesdropper that results from the random distance to the BS.}
We see that the SNR is irrelevant when using LS channel estimation---the eavesdropper dominates the receive signal.
For each 
SNR, the eavesdropper achieves a negative advantage (in dB) in around $75\%$ of the cases, 
and the median advantage is $\delta=-16$\,dB. 
The performance of VILLAIN is far superior and increases with SNR: 
Already for a $0$\,dB SNR, VILLAIN achieves a positive advantage (in dB) in $94\%$ of cases and a median advantage of $\delta=+26$\,dB. 
For an SNR of $15$\,dB and of $30$\,dB, VILLAIN achieves a positive advantage (in dB) in more than $99\%$ of cases, and 
a median advantage of $\delta=+51$\,dB and $\delta=+68$\,dB, respectively.
These results show that VILLAIN succeeds in mitigating active eavesdroppers also in the presence of noise.

\section*{Acknowledgment}
The authors thank A. Stutz for comments and suggestions. 
\appendices
\section{Proofs}

\subsection{Proof of \fref{prop:simo_ortho_guarantee}}
If $\hat\bmh=\bP\tilde\bmh$ for some $\tilde\bmh\in\opC^B$ and
$\bmw = \sqrt{P}\hat\bmh^\ast/\|\hat\bmh\|_2$, then we can write
\begin{align}
	\|\hat\bmh\|_2\tp{\bmj}\bmw &= \sqrt{P} \tp{\bmj}(\bP\tilde\bmh)^\ast
	= \sqrt{P} \tp{\bmj}\bP^\ast \tilde\bmh^\ast 
	\\
	&\stackrel{\text{(a)}}{=} \sqrt{P} \tp{\bmj}\tp{\bP} \tilde\bmh^\ast
	= \sqrt{P} \tp{(\bP\bmj)} \tilde\bmh^\ast 
	\stackrel{\text{(b)}}{=} \tp{\mathbf{0}} \tilde\bmh^\ast
	= 0,
\end{align}
where (a) follows because $\bP$
is an orthogonal projection and so $\herm{\bP}=\bP$
and (b) follows because $\bP\bmj=\mathbf{0}$.
From this, the result follows by dividing both sides by $\|\hat\bmh\|_2\neq0$.	
\hfill$\blacksquare$

\subsection{Proof of \fref{prop:solution_simo_pilot_problem}}
For given $\tilde\bP$, the problem in \eqref{eq:simo_pilot_problem} is quadratic in $\tilde\bmh$, and is minimized by
$\tilde\bmh \!=\! \tilde\bP\bY_T\pinv{(\tp{\bms_T})}$.
Plugging this back into \eqref{eq:simo_pilot_problem} gives
\begin{align}
	\min_{\tilde\bP \,\in\, \mathscr{G}_{B-I}(\opC^B)} \|\tilde\bP \bY_T(\bI_T-\pinv{(\tp{\bms_T})}\tp{\bms_T})\|_F^2,
\end{align}	
which is minimized by $\tilde\bP = \bI_B - \bmu \herm{\bmu}$, where $\bmu\in \opC^{B}$ is the left singular vector 
corresponding to the largest eigenvalue of $\bY_T(\bI_T-\pinv{(\tp{\bms_T})}\tp{\bms_T})$ \cite{eckart1936approximation}.
By plugging this value for $\tilde\bP$ back into $\tilde\bmh = \tilde\bP\bY_T\pinv{(\tp{\bms_T})}$, the result follows. 
\hfill$\blacksquare$

\subsection{Proof of \fref{thm:best_gain}}
Since we assume $\mathsf{N}_{\text{bs}}=0$, we have $\bY_T = \bmh\tp{\bms_T} + \bmj\tp{\bmz}$. 
By \fref{prop:solution_simo_pilot_problem}, the optimal 
$\tilde\bP$ equals $\bI_B - \bmu\herm{\bmu}$, where $\bmu$ is the~left singular vector corresponding to the largest 
singular value of 
\begin{align}
	&\bY_T(\bI_T-\pinv{(\tp{\bms_T})}\tp{\bms_T}) \\
	&= (\bmh\tp{\bms_T} + \bmj\tp{\bmz}) (\bI_T-\pinv{(\tp{\bms_T})}\tp{\bms_T}) \\
	&= \bmh \underbrace{\tp{\bms_T}(\bI_T-\pinv{(\tp{\bms_T})}\tp{\bms_T})}_{=\mathbf{0}} + \bmj\tp{\bmz} (\bI_T-\pinv{(\tp{\bms_T})}\tp{\bms_T}) \\
	&= \bmj\tp{\bmz} (\bI_T-\pinv{(\tp{\bms_T})}\tp{\bms_T}). \label{eq:last_line}
\end{align}
In \eqref{eq:last_line}, $\tp{\bmz} (\bI_T-\pinv{(\tp{\bms_T})}$ is the orthogonal projection of $\bmz$ onto $\text{span}(\bms_T)^\orth$, 
which is distinct from zero if $\bmz\notin \text{span}(\bms_T)$.
Under our assumptions that $T\!>\!1$, that $\bms_T\!\stackrel{\textnormal{i.i.d.}}{\sim}\!\setC\setN(0,1)$ 
and \mbox{$\bmz\!\neq\!\mathbf{0}$} are independent of each other, we have $\bmz\notin \text{span}(\bms_T)$ almost surely.
The rest of the proof is conditioned on this almost sure event. 
It follows that the subspace spanned by $\bmu$ contains $\bmj$, in which case the optimal $\tilde\bP$ 
can be written as
\begin{align}
	\tilde\bP=\bI_B - \bmu\herm{\bmu} = \bI_B - \bmj\pinv{\bmj}.
\end{align}
By \fref{prop:solution_simo_pilot_problem}, the channel estimate by VILLAIN is therefore 
\begin{align}
	\hat\bmh &= (\bI_B - \bmj\pinv{\bmj})\bY_T\pinv{(\tp{\bms_T})} \\
	&= (\bI_B - \bmj\pinv{\bmj})(\bmh\tp{\bms_T} + \bmj\tp{\bmz})\pinv{(\tp{\bms_T})} \\
	&= (\bI_B - \bmj\pinv{\bmj})\bmh\underbrace{\tp{\bms_T}\pinv{(\tp{\bms_T})}}_{=1} 
		+ \underbrace{(\bI_B - \bmj\pinv{\bmj}) \bmj}_{=\mathbf{0}} \tp{\bmz} \pinv{(\tp{\bms_T})} \\ 
	&= (\bI_B - \bmj\pinv{\bmj})\bmh. 
\end{align}
When using MRT precoding, we get the precoding vector 
\begin{align}
	\bmw = (\sqrt{P}/\|\bP\bmh\|_2) (\bP\bmh)^{\ast}, \label{eq:mrt_solution}
\end{align}
where we define $\bP\triangleq \bI_B - \bmj\pinv{\bmj}$ for the remainder of the proof. 

We now show that the solution to the  optimization problem in \eqref{eq:opt_problem_ue_rec_power} 
coincides with \eqref{eq:mrt_solution}. 
The set of vectors $\tilde\bmw$ that satisfy the constraint $\tp{\bmj}\tilde\bmw=0$ 
is simply the image of $\bP^\ast$ and can be rewritten 
by substituting $\tilde\bmw$ with~$\bP^\ast\bar\bmw$:
\begin{align}
	\{\tilde\bmw: \tilde\bmw\in\opC^B,\tp{\bmj}\tilde\bmw=0\} = \{\bP^\ast\bar\bmw: \bar\bmw\in\opC^B\}.
\end{align}
The problem in \eqref{eq:opt_problem_ue_rec_power} can therefore be reformulated as
\begin{align}
	\max_{\bar\bmw\in\opC^B:\|\bP^\ast\bar\bmw\|_2^2\leq P} |\tp{\bmh}\bP^\ast\bar\bmw|^2 
	&= \max_{\|\bP^\ast\bar\bmw\|_2^2\leq P} |\tp{\bmh}\tp{\bP}\bar\bmw|^2 \\
	&= \max_{\|\bP^\ast\bar\bmw\|_2^2\leq P} |\tp{(\bP\bmh)}\bar\bmw|^2, 
\end{align}
which is solved by $\bar\bmw \propto (\hat\bP\bmh)^\ast$. By reinserting $\tilde\bmw=\bP^\ast\bar\bmw$, 
and respecting the power constraint, we get the~$\tilde\bmw$ that solves \eqref{eq:opt_problem_ue_rec_power}: 
\begin{align}
	\tilde\bmw &= \sqrt{P} \frac{ \bP^\ast(\bP\bmh)^\ast}{\|\bP^\ast(\bP\bmh)^\ast\|_2} 
	= \sqrt{P} \frac{(\bP\bmh)^\ast}{\|\bP\bmh\|_2}, \label{eq:opt_w_vector_ue_power}
\end{align}
which coincides with \eqref{eq:mrt_solution}. 
The signal power at the UE is
\begin{align}
	\!\!\! |\tp{\bmh}\bmw|^2 &= P \frac{|\tp{\bmh}(\bP\bmh)^\ast|^2}{\|\bP\bmh\|_2^2} 
	= P \frac{|\herm{\bmh}\bP\bmh|^2}{\|\bP\bmh\|_2^2}
	= P \frac{|\herm{\bmh}\herm{\bP}\bP\bmh|^2}{\|\bP\bmh\|_2^2} \!\! \\
	&= P \|\bP\bmh\|_2^2 
	= P \|(\bI_B-\bmj\pinv{\bmj})\bmh\|_2^2.
\end{align}
This concludes the proof.
\hfill$\blacksquare$

\linespread{1.01}
% Generated by IEEEtran.bst, version: 1.14 (2015/08/26)


% Generated by IEEEtran.bst, version: 1.14 (2015/08/26)
\begin{thebibliography}{10}
\providecommand{\url}[1]{#1}
\csname url@samestyle\endcsname
\providecommand{\newblock}{\relax}
\providecommand{\bibinfo}[2]{#2}
\providecommand{\BIBentrySTDinterwordspacing}{\spaceskip=0pt\relax}
\providecommand{\BIBentryALTinterwordstretchfactor}{4}
\providecommand{\BIBentryALTinterwordspacing}{\spaceskip=\fontdimen2\font plus
\BIBentryALTinterwordstretchfactor\fontdimen3\font minus
  \fontdimen4\font\relax}
\providecommand{\BIBforeignlanguage}[2]{{%
\expandafter\ifx\csname l@#1\endcsname\relax
\typeout{** WARNING: IEEEtran.bst: No hyphenation pattern has been}%
\typeout{** loaded for the language `#1'. Using the pattern for}%
\typeout{** the default language instead.}%
\else
\language=\csname l@#1\endcsname
\fi
#2}}
\providecommand{\BIBdecl}{\relax}
\BIBdecl

\bibitem{whitman2021principles}
M.~E. Whitman and H.~J. Mattord, \emph{Principles of Information
  Security}.\hskip 1em plus 0.5em minus 0.4em\relax Cengage learning, 2021.

\bibitem{bloch2011physical}
M.~Bloch and J.~Barros, \emph{Physical-Layer Security: From Information Theory
  to Security Engineering}.\hskip 1em plus 0.5em minus 0.4em\relax Cambridge
  Univ. Press, 2011.

\bibitem{wu2018survey}
Y.~Wu, A.~Khisti, C.~Xiao, G.~Caire, K.-K. Wong, and X.~Gao, ``A survey of
  physical layer security techniques for {5G} wireless networks and challenges
  ahead,'' \emph{{IEEE} J. Sel. Areas Commun.}, vol.~36, no.~4, pp. 679--695,
  2018.

\bibitem{li2007secret}
Z.~Li, W.~Trappe, and R.~Yates, ``Secret communication via multi-antenna
  transmission,'' in \emph{Proc. 41st Ann. Conf. Inf. Sci. Syst.}, 2007, pp.
  905--910.

\bibitem{shafiee2007achievable}
S.~Shafiee and S.~Ulukus, ``Achievable rates in {Gaussian MISO} channels with
  secrecy constraints,'' in \emph{Proc. IEEE Int. Symp. Inf. Theory (ISIT)},
  2007, pp. 2466--2470.

\bibitem{rezki2011finite}
Z.~Rezki and M.-S. Alouini, ``On the finite-{SNR} diversity-multiplexing
  tradeoff of zero-forcing transmit scheme under secrecy constraint,'' in
  \emph{Proc. IEEE Int. Conf. Commun. (ICC)}, 2011, pp. 1--5.

\bibitem{reboredo2013filter}
H.~Reboredo, J.~Xavier, and M.~R. Rodrigues, ``Filter design with secrecy
  constraints: The {MIMO Gaussian} wiretap channel,'' \emph{{IEEE} Trans.
  Signal Process.}, vol.~61, no.~15, pp. 3799--3814, 2013.

\bibitem{kapetanovic2015physical}
D.~Kapetanovic, G.~Zheng, and F.~Rusek, ``Physical layer security for massive
  {MIMO}: An overview on passive eavesdropping and active attacks,''
  \emph{{IEEE} Commun. Mag.}, vol.~53, no.~6, pp. 21--27, 2015.

\bibitem{bereyhi2018robustness}
A.~Bereyhi, S.~Asaad, R.~R. M{\"u}ller, R.~F. Schaefer, and A.~M. Rabiei, ``On
  robustness of massive {MIMO} systems against passive eavesdropping under
  antenna selection,'' in \emph{Proc. IEEE Global Commun. Conf. (GLOBECOM)},
  2018, pp. 1--7.

\bibitem{bereyhi2019robustness}
A.~Bereyhi, S.~Asaad, R.~R. M{\"u}ller, R.~F. Schaefer, G.~Fischer, and H.~V.
  Poor, ``Robustness of low-complexity massive {MIMO} architectures against
  passive eavesdropping,'' \emph{arXiv preprint arXiv:1912.02444}, 2019.

\bibitem{zhou2012pilot}
X.~Zhou, B.~Maham, and A.~Hj{\o}rungnes, ``Pilot contamination for active
  eavesdropping,'' \emph{{IEEE} Trans. Wireless Commun.}, vol.~11, no.~3, pp.
  903--907, 2012.

\bibitem{wu2016secure}
Y.~Wu, R.~Schober, D.~W.~K. Ng, C.~Xiao, and G.~Caire, ``Secure massive {MIMO}
  transmission with an active eavesdropper,'' \emph{{IEEE} Trans. Inf. Theory},
  vol.~62, no.~7, pp. 3880--3900, 2016.

\bibitem{bereyhi2020secure}
A.~Bereyhi, S.~Asaad, R.~R. M{\"u}ller, R.~F. Schaefer, and H.~V. Poor,
  ``Secure transmission in {IRS-assisted MIMO} systems with active
  eavesdroppers,'' in \emph{Proc. Asilomar Conf. Signals, Syst., Comput.},
  2020, pp. 718--725.

\bibitem{li2017mimo}
L.~Li, A.~P. Petropulu, and Z.~Chen, ``{MIMO} secret communications against an
  active eavesdropper,'' \emph{{IEEE} Trans. Inf. Forensics Security}, vol.~12,
  no.~10, pp. 2387--2401, 2017.

\bibitem{cho2020zero}
S.~Cho, G.~Chen, and J.~P. Coon, ``Zero-forcing beamforming for active and
  passive eavesdropper mitigation in visible light communication~systems,''
  \emph{{IEEE} Trans. Inf. Forensics Security}, vol.~16, pp. 1495--1505, 2021.

\bibitem{si2020cooperative}
J.~Si, Z.~Cheng, Z.~Li, J.~Cheng, H.-M. Wang, and N.~Al-Dhahir, ``Cooper-ative
  jamming for secure transmission with both active and
  passive~eaves-droppers,'' \emph{{IEEE} Trans. Commun.}, vol.~68, no.~9, pp.
  5764--5777, 2020.

\bibitem{cho2021cooperative}
S.~Cho, G.~Chen, and J.~P. Coon, ``Cooperative beamforming and jamming for
  secure vlc system in the presence of active and passive eavesdroppers,''
  \emph{IEEE Trans. Green Commun. Netw.}, vol.~5, no.~4, pp. 1988--1998, 2021.

\bibitem{zhou2022caching}
Y.~Zhou, P.~L. Yeoh, C.~Pan, K.~Wang, Z.~Ma, B.~Vucetic, and Y.~Li, ``Caching
  and {UAV} friendly jamming for secure communications with active
  eavesdropping attacks,'' vol.~71, no.~10, pp. 11\,251--11\,256, 2022.

\bibitem{yu2020robust}
X.~Yu, D.~Xu, Y.~Sun, D.~W.~K. Ng, and R.~Schober, ``Robust and secure wireless
  communications via intelligent reflecting surfaces,'' \emph{{IEEE} J. Sel.
  Areas Commun.}, vol.~38, no.~11, pp. 2637--2652, 2020.

\bibitem{jia2023star}
H.~Jia, L.~Ma, and S.~Valaee, ``{STAR-RIS} enabled downlink secure {NOMA}
  network under imperfect {CSI} of eavesdroppers,'' \emph{{IEEE} Commun.
  Lett.}, vol.~27, no.~3, pp. 802--806, 2023.

\bibitem{jacobsson16d}
S.~Jacobsson, G.~Durisi, M.~Coldrey, T.~Goldstein, and C.~Studer, ``Nonlinear
  1-bit precoding for massive {MU-MIMO} with higher-order modulation,'' in
  \emph{Proc. Asilomar Conf. Signals, Syst., Comput.}, Pacific Grove, CA, USA,
  2016, pp. 763--767.

\bibitem{khisti2010secure}
A.~Khisti and G.~W. Wornell, ``Secure transmission with multiple
  antennas--{Part I}: The {MISOME} wiretap channel,'' \emph{{IEEE} Trans. Inf.
  Theory}, vol.~56, no.~7, pp. 3088--3104, 2010.

\bibitem{jaeckel2014quadriga}
S.~Jaeckel, L.~Raschkowski, K.~B{\"o}rner, and L.~Thiele, ``{QuaDRiGa}: A {3-D}
  multi-cell channel model with time evolution for enabling virtual~field
  trials,'' \emph{IEEE Trans.\! Antennas Propag.}, vol. \!62, no. \!6, pp.
  \!3242--3256, 2014.

\bibitem{3gpp22a}
3GPP, ``{3GPP TR 38.901},'' Mar. 2022, version 17.0.0.

\bibitem{eckart1936approximation}
C.~Eckart and G.~Young, ``The approximation of one matrix by another of lower
  rank,'' \emph{Psychometrika}, vol.~1, no.~3, pp. 211--218, 1936.

\end{thebibliography}
\end{document}